\documentclass[12pt]{article}
\usepackage{latexsym}
\usepackage{amssymb}
\usepackage[dvips]{graphicx}
\usepackage{latexsym}
\newcommand{\C}{\mathbb{C}}
\newcommand{\Th}{\Theta}
\newcommand{\CP}{\mathbb{CP}}

\newcommand{\spp}{\mathbb{S}}
\newcommand{\R}{\mathbb{R}}

\def\p{\partial}
\def\t{\tilde}
\def\Sm{\Sigma}
\def\l{\lambda}

\renewcommand{\d}{\mathrm{d}}
\def\be{\begin{equation}}
\def\ee{\end{equation}}
\def\theequation{\thesection.\arabic{equation}}
\begin{document}
\pagestyle{plain}

\title{\vskip -70pt
\begin{flushright}
{\normalsize DAMTP-2008-27} \\
\end{flushright}
\vskip 80pt
{\bf Interpolating Dispersionless Integrable System}
\vskip 20pt}
\author{Maciej Dunajski\thanks{email m.dunajski@damtp.cam.ac.uk}\\[15pt]
{\sl Department of Applied Mathematics and Theoretical Physics} \\[5pt]
{\sl University of Cambridge} \\[5pt]
{\sl Wilberforce Road, Cambridge CB3 0WA, UK} 
}
\date{} 
\maketitle
\begin{abstract}
We introduce a dispersionless integrable system
which interpolates between the dispersionless Kadomtsev--Petviashvili
equation
and the hyper--CR equation. The interpolating system
arises as a symmetry reduction of the anti--self--dual
Einstein equations in (2, 2) signature by a conformal Killing
vector whose self--dual derivative is null.  It also arises 
as a special case of the Manakov--Santini integrable system.
We discuss the corresponding Einstein--Weyl structures.
\end{abstract}
\newpage
\section{Introduction}
\setcounter{equation}{0}
It has been known for more than 20 years that many integrable systems
admitting soliton solutions arise as symmetry reductions of anti--self--dual
Yang Mills (ASDYM) equations in four dimensions 
\cite{Wa85}. The Riemann--Hilbert
factorisation problem underlies this approach to integrability: it
appears in classical solution generating techniques like
dressing transformations \cite{NMPZ}, as well as in the twistor treatment of 
ASDYM \cite{Wa77}. 

The dispersionless integrable systems in 2+1 dimensions do not fit into this 
framework: they do not admit soliton solutions and there is no
associated Riemann--Hilbert problem where the corresponding Lie group is
finite dimensional. These systems can nevertheless be described in terms
of anti--self--duality (ASD) conditions on a four--dimensional conformal 
structure. In this case the `unknown' in the equations  
is not a gauge field, but rather a metric (up to scale) on some 
four--manifold \cite{Pe76}. 
This makes the dispersionless systems more geometric
than their solitonic cousins. This point of view
may be of deep significance in description of shock formations:
a recent beautiful analysis of Manakov and Santini \cite{MSshock} deduced the
gradient catastrophe of the localised solutions to the dKP equation
using the inverse scattering transform.
It may be however,  that this catastrophe is 
only an artifact of a chosen coordinate system, and the underlying conformal structure is regular, but needs to be covered 
by more than one coordinate patch. It remains to be seen whether this
is indeed the case.

To classify existing 2+1 dispersionless integrable systems, and
perhaps discover some new ones one needs to classify
the symmetry reductions of the conformal ASD equations. If the 
Ricci--flat condition is imposed on top of the anti--self--duality,
the work of Pleba\'nski \cite{Pl75} implies the existence
of a local coordinate system $(X, Y, W, Z)$  and a function $\Theta$ 
on an open set  ${\cal M}\subset\R^4$
such
than any ASD Ricci--flat metric is locally of the form
\be
\label{Plebanm}
g=2(\d Z\d Y +\d W\d X -\Th_{XX}\d Z^2-\Th_{YY}\d W^2+2\Th_{XY}\d W\d Z),
\ee
where $\Theta_{XY}=\p_X\p_Y\Theta$ etc, and $\Theta$ satisfies the second heavenly equation
\be
\label{secondeq}
\Th_{ZY}+\Th_{WX}+\Th_{XX}\Th_{YY}-\Th_{XY}^2=0.
\ee
This metric has signature $(+ + - -)$ but this is precisely what need:
Given a non--null symmetry, 
the conformal structure on a space of orbits will have signature
$(2, 1)$ and will be described by a hyperbolic integrable equation. 
We are thus led to study (\ref{Plebanm}), 
(\ref{secondeq}) subject to the existence of
a conformal Killing vector $K$
\[
{\cal L}_Kg=cg,\qquad g(K, K)\neq 0
\]
where $c$ is some function. 
The classification of reductions is based on studying the action of $K$
on the bundle of self--dual two--forms over 
${\cal M}$. Assume that ${\cal M}$ is oriented and
recall that given a metric of $(+ + - -)$ signature 
on ${\cal M}$, the
Hodge $\ast$ operator is an  involution on two--forms and induces a decomposition 
\[
\Lambda^{2} = \Lambda_{+}^{2} \oplus \Lambda_{-}^{2}
\]
of two--forms  into self--dual  and anti--self--dual components.
Moreover there exist real two-dimensional vector bundles  $\spp$ and  $\spp'$  (called spin bundles) over ${\cal M}$
such that $T {\cal M}\cong {\spp}\otimes {\spp'}$ and
$
\Lambda^{2}_{+} \cong \spp \odot \spp.
$
Therefore the self--dual derivative of (the one--form
metric dual to)
$K$ 
\[
\d K_+:=\frac{1}{2}(\d K+\ast \d K)
\]
corresponds to a symmetric 2 by 2 matrix $\phi$
explicitly given by
\[
\d K_+=\phi_{AB}\Sigma^{AB}, \qquad A, B=0, 1,
\]
where the self--dual two forms\footnote{The ASD Ricci--flat equations
are equivalent to
\[
\d \Sigma^{AB}=0, \quad \Sigma^{(AB}\wedge\Sigma^{CD)}=0, \qquad A, B, C, D =0,1
\]
and the second heavenly equation 
(\ref{secondeq}) arises by using the Darboux
theorem to introduce coordinates such
that
\[
\Sm^{00}=\d W\wedge \d Z,\quad \Sm^{01}=\d W\wedge 
\d X+\d Z\wedge\d Y,
\]
and deducing the existence of the function $\Theta$
from the remaining conditions on $\Sigma^{AB}$.}
$(\Sigma^{00}, \Sigma^{01}, \Sigma^{11})$ span $\Lambda^2_+$.
The rank of the matrix $\phi$ does not depend on the choice
of the basis $\Sigma^{AB}$ and the
classification is based on this rank. In the Riemannian signature this can only be 0 or 2  (the former
case is called the tri--holomorphic symmetry),
but in the $(+ + - -)$ signature we can also have
rank$(\phi)=1$, in which case $(\d K)_+\wedge
(\d K)_+=0$ so the self--dual derivative of $K$
is null. 

This classification programme has
almost been completed and one aim of this paper is to remove the question mark from the following
table summarising the reductions
of the second heavenly equation (\ref{secondeq})
\vskip5pt
\centerline{ 
\noindent \begin{tabular}{c|c|c|}
&$c=0$ & $c\neq0 $\\
\hline rank$(\phi)=0$ &2+1 Linear wave equation  \cite{GH78, FP79}& 
Hyper--CR equation \cite{thesis, D04} \\
\hline rank$(\phi)=1$ &
Dispersionless KP equation \cite{DMT01}& 
{\bf ?}\\
\hline rank$(\phi)=2$ &
SU$(\infty)$ Toda equation \cite{FP79} & Integrable equation studied in \cite{DT99}\\
\hline
\end{tabular}}
\vskip5pt
The case where rank$(\phi)=1$ and $c\neq 0$ 
has not yet been investigated,
and the resulting integrable system is the subject of the present paper.
This system is given by
\be
\label{uw_eq}
u_y+w_x=0, \qquad u_t+w_y-c(uw_x-wu_x)+buu_x=0,
\ee
where $b$ and $c$ are constants and $u, w$ are smooth functions of
$(x,y,t)$. We propose to call (\ref{uw_eq}) an interpolating system as it
contains two well known dispersionless equations as the limiting cases:
Setting $b=0, c=-1$ gives the hyper--CR equation 
\cite{thesis, Pa03, FK04, D04, MaSh02,  MS07, OR2} 
and setting $c=0, b=1$  gives the dKP equation. In fact
one constant can always be eliminated from (\ref{uw_eq}) 
by redefining the coordinates and
it is only the ratio of $b/c$ which 
remains\footnote{Jenya Ferapontov has pointed out that 
if $b\neq 0$ the hydrodynamic reductions of (\ref{uw_eq})
coincide with the hydrodynamic reductions of the dKP equation,
despite the fact that   
dKP and (\ref{uw_eq}) are not point or contact equivalent unless $c=0$.
}.  We prefer
to keep both constants as it makes various limits
more transparent.

In the next Section we shall give the dispersionless Lax pair
for (\ref{uw_eq}). The Lie group underlying the Lax formulation is 
Diff $(\Sigma^2)$ -- an infinite--dimensional  group
of diffeomorphisms of some two--dimensional manifold $\Sigma^2$.
The interpolating system 
corresponds to Lorentzian Einstein--Weyl structures in three dimensions.
This gives an intrinsic geometric interpretation without
the need of going to four--dimensions. This will be
described in Section \ref{section3}. 
In Section \ref{section4} we shall show that (\ref{uw_eq})
is a special case of the 
Manakov--Santini integrable system \cite{MS06,MS07}
and find the Manakov--Santini Einstein--Weyl structure.

The explicit reduction of the second heavenly equation (\ref{secondeq})
to the interpolating system (\ref{uw_eq})  will be presented in the Appendix.
In particular we shall show that the most general $(+ + - -)$
ASD Ricci flat metric with a conformal Killing vector
whose self--dual derivative is null
is of the form
\be
\label{metric}
g=e^{c\phi}(Vh -V^{-1}(\d \phi +A)^2),
\ee
where $\phi$ parametrises the orbits of the conformal Killing 
vector $K=\p/\p\phi$ and
\[
h=(\d y+cu\d t)^2-4(\d x+cu\d y-(cw+bu)\d t)\d t,\quad
A=-\frac{1}{2}u_x\d y+\Big(\frac{c}{2}uu_{x}-u_{y}\Big)\d t,\quad
V=\frac{1}{2}u_x.
\]
This reduction explains the origin of the two parameters
$(b, c)$ in (\ref{uw_eq})  as in this case the conformal symmetry is
\[
K=c\times\mbox{(dilatation)}+b\times\mbox{
(rotation with null SD derivative).}
\]
\section{Lax pair and Diff$(\Sigma^2)$ hierarchies}
\setcounter{equation}{0}
The system (\ref{uw_eq}) admits a dispersionless Lax pair 
\be
\label{interp_lax}
L_0=\frac{\p}{\p t}+(cw+bu-\l cu-\l^2)\frac{\p}{\p x}+
b(w_x-\l u_x)\frac{\p}{\p \l}, \quad
L_1=\frac{\p}{\p y}-(cu +\l)\frac{\p}{\p x}-bu_x\frac{\p}{\p\l}
\ee
with a spectral parameter $\lambda\in \CP^1$.
The overdetermined system of linear equations $L_0\Phi=L_1\Phi=0$, where
$\Phi=\Phi(x, y, t, \l)$ admits solutions because equations 
(\ref{uw_eq}) are equivalent to  $[L_0, L_1]=0$.

In general, consider the vector fields of the form
\be
\label{distribution_L}
L_i=\frac{\p}{\p t_i}+A_i\frac{\p}{\p x}+B_i\frac{\p}{\p \lambda},
\ee
where $A_i, B_i$ are polynomials in
$\lambda$ with coefficients 
depending on  $(t^0=x, t^i)$.
The flows of  the Diff$(\Sigma^2)$ hierarchy are defined by
\be
\label{s2hierachy}
[L_i, L_j]=0,\qquad i,j=1, \dots, n.
\ee
To achieve a dual formulation, generalising Krichever's approach
to dispersionless integrable systems \cite{Kri94},
complexify the hierarchy (so that $(t^0, t^i)\in \C^{n+1}$)
and define a two--form $\Omega$ on $\C^{n+1}\times\CP^1$ by
\[
\Omega(X, Y)=\d t_1\wedge \dots \wedge \d t_n\wedge \d x\wedge \d \lambda(
L_1, \dots, L_n, X, Y)
\]
so that
\[
\Omega=\d x\wedge\d\lambda+\sum_i (A_i\d \lambda-~B_i\d x)\wedge\d t_i
+\sum_{i, j} (A_iB_j-B_jA_i)\d t_i\wedge \d t_j.
\]
The two--form $\Omega$ is simple and 
satisfies the Frobenius integrability  conditions
\[
\Omega\wedge\Omega=0, \qquad\d \Omega =\Omega\wedge\beta
\]
for some one--form $\beta$. We recover various dispersionless
hierarchies as special cases of this formulation
\begin{itemize}
\item {\bf SDiff$(\Sigma^2)$ hierarchy}. 
The group Diff$(\Sigma^2)$ reduces to SDiff$(\Sigma^2)$
generated by Hamiltonian vector fields. The corresponding Lie algebra is 
homomorphic with the Poisson bracket algebra.
 The vector fields $L_i$ preserve the two--form 
$\d x\wedge \d \lambda$ and
\[
A_i=\frac{\p H_i}{\p \lambda}, \qquad B_i=-\frac{\p H_i}{\p x}.
\]
The two--form is given by 
$\Omega =\d x\wedge\d\lambda+\sum_i \d H_i\wedge \d t_i$
(where we have used (\ref{s2hierachy})), 
and the  SDiff$(\Sigma^2)$ hierarchy is given by 
\[
\Omega\wedge\Omega=0, \qquad \d\Omega=0,
\]
which is the original Krichever's formulation \cite{Kri94}.
The Darboux theorem implies the existence of functions $P, Q$ such that
$\Omega =\d P\wedge \d Q$. These two functions are local coordinates
on the twistor space, which is a quotient of $\C^{n+1}\times\CP^1$
by the integrable distribution $\{ L_i \}$.
Both the dKP and SU$(\infty)$ Toda hierarchies fit into this category
\cite{Kri94, takasaki, DMT01}.
\item {\bf Diff$(S^1)$ hierarchy}. The group Diff$(\Sigma^2)$ reduces to Diff$(S^1)$,
where $\Sigma^2=TS^1$. This case corresponds to
\[B_i=0.\] The underlying Lie algebra is that of a Wronskian with a Lie bracket
$<f, g>=f_xg-fg_x$.
In this case $\Omega = e\wedge \d \lambda$, where 
$e=\d x-\sum_iA_i\d t_i$. This one--form is integrable in the Frobenius sense
\[
e\wedge \d e=0,
\]
where here $\d$ keeps $\lambda=$const. The twistor space fibres holomorphically
over $\CP^1$. The hyper-CR hierarchy \cite{D04} and the universal hierarchy 
\cite{MaSh02} are of this type.
\end{itemize}
\section{Interpolating Einstein--Weyl structure}
\label{section3}
\setcounter{equation}{0}
A three--dimensional Lorentzian Weyl structure  $(M, D, [h])$ consists 
of a 3--manifold $M$, a
torsion-free connection $D$ and a
conformal metric $[h]$ of Lorentzian signature such that the
null geodesics of $[h]$ are also geodesics for $D$.  This condition is
equivalent to 
\[
Dh=\omega\otimes h
\]
for some one form $\omega$. Here $h$ is a representative metric in
the conformal class. If we change this representative by
$h\longrightarrow \gamma^2 h$, then $\omega\longrightarrow
\omega+2\d\ln{\gamma}$. A Weyl structure is called Einstein--Weyl
if the conformally invariant equations
\be
\label{EW_equations}
R_{(ab)}=\Lambda h_{ab}, \qquad a, b, =1, \dots, 3
\ee
hold for some function $\Lambda$. Here $R_{(ab)}$ is the symmetrised Ricci tensor of $D$ and
$h_{ab}$ is a representative metric in a conformal class $[h]$.
In practice the Einstein--Weyl structure is given by specifying the metric
$h\in [h]$, and the one--form $\omega$ which measures
the difference between the Weyl connection $D$ and the Levi--Civita connection of 
$h$.

The three--dimensional Einstein--Weyl condition is a dispersionless
integrable system \cite{DMT01}: Let $Z, W,  \widetilde{W}$ be independent
vector fields on $M$ such that a contravariant metric in $[h]$ is  
\[
h=h^{ab}\frac{\p}{\p x^a}\otimes\frac{\p}{\p x^b}=
Z\otimes Z-2(W\otimes\widetilde{W}+\widetilde{W}\otimes W).
\]
Then there exists a connection $D$ such that $(M, [h], D)$ is Einstein--Weyl
if the dispersionless Lax pair
\be
\label{EW_lax}
L_0=W-\lambda Z+f_0\frac{\p}{\p\lambda},\quad
L_1=Z-\lambda \widetilde{W}+f_1\frac{\p}{\p\lambda},
\ee
satisfies the integrability condition
\[
[L_0, L_1]=0 \quad\mbox{modulo}\quad\,L_0, L_1
\]
for some functions $(f_0, f_1)$ which are cubic polynomials in
$\lambda\in \CP^1$. Conversely, every Einstein--Weyl structure arises from some Lax pair (\ref{EW_lax}).
The corresponding 
one form $\omega$ can be read off from the Levi--Civita connection of
$h$ and the coefficients of  $(f_0, f_1)$.
This Lax formulation has a geometric
origin which goes back to E. Cartan \cite{CartanEW}
\begin{itemize}
\item
Einstein--Weyl condition is equivalent to the existence
of a two parameter family of totally geodesic null surfaces in $M$.
\end{itemize}
Comparing this with the
Lax pair (\ref{interp_lax}) for the interpolating system, and
taking linear combinations to put (\ref{interp_lax}) in the form
(\ref{EW_lax}) we find the corresponding Einstein--Weyl structure to be
\begin{eqnarray}
\label{EW_interpol}
h&=&(\d y+cu\d t)^2-4(\d x+cu\d y-(cw+bu)\d t)\d t,\\
\omega&=&-cu_x\d y+(4bu_x+c^2uu_x-2cu_y)\d t.\nonumber
\end{eqnarray}
Taking the limits we recover various known Einstein--Weyl structures
from (\ref{EW_interpol}). These structures 
can be characterised by the properties  null 
shear-free and geodesic congruence $\d t$ 
(this  elegant  framework is described, in the Riemannian case, in
\cite{CP})
\begin{itemize}
\item  Setting $b=0, c=-1$ gives the hyper--CR Einstein--Weyl structure
\cite{D04}. The  congruence is divergence--free. This is 
a Lorentzian analogue Einstein--Weyl structures studied in
\cite{GT98, CP, DT99}.
\item  Setting $c=0, b=1$  gives the dKP Einstein--Weyl spaces \cite{DMT01}. 
The congruence $\d t$ is now twist-free, and the dual vector
$\p/\p x$ is parallel with a weight $-1/2$ with respect to the Weyl
connection.
\end{itemize}
We remark that the Einstein--Weyl (\ref{EW_interpol}) structure 
can also be read off from the ASD Ricci--flat metric (\ref{metric}) using the Jones--Tod correspondence \cite{JT85}.
\section{The Manakov--Santini system}
\label{section4}
\setcounter{equation}{0}
The system (\ref{uw_eq}) is a special case of the Manakov--Santini 
system \cite{MS06,MS07}
\begin{eqnarray}
\label{ZM}
&&U_{xt}-U_{yy}+(UU_x)_x+V_xU_{xy}-V_yU_{xx}=0\\
&&V_{xt}-V_{yy}+UV_{xx}+V_xV_{xy}-V_yV_{xx}=0\nonumber,
\end{eqnarray}
where $U=U(x,y,t), V=V(x, y, t)$.
To see it, notice that the first equation in (\ref{uw_eq}) 
implies the existence of $v(x, y)$ such that $u=v_x, w=-v_y$, and
$v$ satisfies 
\be
\label{eq1}
v_{xt}-v_{yy}+c(v_xv_{xy}-v_yv_{xx})+buv_{xx}=0.
\ee
Differentiating the second equation in  (\ref{uw_eq}), and eliminating
$w$ yields
\be
\label{eq2}
u_{xt}-u_{yy}-c(v_yu_{xx}-v_xu_{xy})+b(uu_x)_x=0.
\ee
Now assume the generic case when the 
constants $c, b$ are non zero, and set $U=bu, V=cv$. Then
the system (\ref{eq1}) and (\ref{eq2}) is equivalent to 
(\ref{ZM}) with an additional constraint
\be
\label{constraint}
cU-bV_x=0.
\ee
The Manakov--Santini system is more general than (\ref{uw_eq}):
Regarding the second equation in (\ref{ZM}) as the definition of $U$,
and substituting $U$ to the first equation in (\ref{ZM}) yields
a fourth order scalar PDE for $V$. Thus the naive counting suggest
that the general solution to (\ref{ZM}) depends on 4 functions of 2
variables (a caution is needed as the resulting PDE is not in the 
Cauchy--Kowalewska form). 
In the special case when the constraint (\ref{constraint}) holds 
both equations in (\ref{ZM}) reduce to a single second
order PDE for $V$. The solution depends on two functions of two variables, and
a constant (the ratio $b/c$).

The Manakov--Santini system also corresponds to an Einstein--Weyl structure
\begin{eqnarray}
\label{Manakov_Santini_EW}
h&=&(\d y-V_x  \d t)^2-4(\d x-(U-V_y)\d t)\d t,\\
\omega&=&-V_{xx}\d y+(4U_x-2V_{xy}+V_xV_{xx})\d t.\nonumber
\end{eqnarray}
To verify it set  $x^a=(y,x,t)$. The $(11), (12), (22), (23)$  components
of the Einstein--Weyl equations hold identically. The $(13)$ component vanishes if the second equation in
(\ref{ZM}) holds, and finally the
$(33)$ component vanishes if
both equations in $(\ref{ZM})$ are satisfied.

Conversely,  consider the general conformal structure in $(2+1)$ dimensions
given in local coordinates $x^a$ by a representative metric
\[
h={\left (
\begin{array}{ccc}
h_{11}&h_{12}&h_{13} \\
h_{12}&h_{22}&h_{23}\\
h_{13}&h_{23}&h_{33}
\end{array}
\right ).
}
\]
Using the diffeomorphism freedom, and 
the conformal rescaling we can impose
four constraint on six functions
$h_{ab}(x^c)$ as long as
the resulting quadratic form is
non--degenerate. We choose to set
$h_{11}=h_{12}=0, h_{13}=-2, 
h_{23}=-A, h_{33}=A^2+4B$,
where $A$ and $B$ are some functions
of $(x, y, t)$, so that\footnote{This is analogous to the existence
of orthogonal coordinates in three dimensions. The proof is relatively straightforward
in the real--analytic category, and more subtle in the smooth category.}
\[
h=(\d y-A\d t)^2-4(\d x-B\d t)\d t.
\]
Now given $A, B$ we can always find two functions $U, V$ such that
$A=V_x, B=U-V_y$ so that the metric is in the form (\ref{Manakov_Santini_EW})

Now we find the corresponding dual basis, 
and construct the Lax pair (\ref{EW_lax}). Before imposing the integrability conditions
it is convenient to take a linear combination of the vectors in this distribution, so that the resulting
pair of vectors commutes exactly.
This yields
\[
L_0=\p_y-(\lambda+A)\p_x+f_0\p_\l,
\quad L_1=\p_t-(\l^2+\l A-B)\p_x
+f_1\p_\l,
\]
where the polynomials $f_0$ and $f_1$
are respectively  cubic and quartic in $\lambda$. There is some additional
freedom which will preserve the above
form of the Lax pair. We can translate
the fibres by 
$\l\rightarrow \l+
\kappa(x, y, t)$. We use this freedom
to set the linear term in $f_0$ to zero. It is possible that
some further coordinate freedom can used to set the quadratic term
in $f_0$ to zero
 which would imply that
$\omega$ is  given by (\ref{Manakov_Santini_EW}). This would imply that
that every Einstein--Weyl structure is equivalent to the Manakov--Santini 
EW  structure (\ref{Manakov_Santini_EW}). So far we have been  unable to find
the right transformation, and we need to impose $\p_\lambda^2 f_0=0$ 
as an additional condition. Then  the integrability condition $[L_0, L_1]=0$
implies (\ref{ZM}).

\section*{Acknowledgements}
I thank David Calderbank, 
Jenya Ferapontov  and Luis Mart\'inez--Alonso
for useful comments.
This work is partly supported by  the European network Methods of Integrable Systems, Geometry and 
Applied Mathematics (MISGAM).
\section*{Appendix. Reduction of the second heavenly equation}
\setcounter{equation}{0}
\appendix
\def\theequation{\thesection{A}\arabic{equation}}
We shall prove that
the system (\ref{uw_eq}) arises as the most general symmetry reduction of the second heavenly equation
(\ref{secondeq}) by a conformal Killing vector
with null self-dual derivative. 

 Let $\Th=\Th(W, Z, X, Y)$
satisfy  the second heavenly equation 
and let the corresponding metric be given by (\ref{Plebanm}).
Let $K$ be a conformal Killing vector for 
(\ref{Plebanm}). 
Using Penrose's two-component spinor formalism  
we can show that the conformal Killing equations and the Ricci identity imply that $\nabla_{AA'}{K^A}_{B'}$ is covariantly constant, or otherwise $g$ is of Petrov--Penrose
type $N$ an can be found explicitly. In the spin
frame of the heavenly metric (\ref{Plebanm}) the connection on the 
spin bundle $\spp$ vanishes,  so $\nabla_{A'A}{K^{A'}}_{B}$
is in fact constant. 
We are interested in the case where the self--dual 
derivative
$\phi_{AB}=\nabla_{A'(A}{K^{A'}}_{B)}$
is of rank one. Therefore we need to integrate the linear system
\[
\nabla_{A'A}{K^{A'}}_{B}=\left (
\begin{array}{cc}
0&c\\
-c &b
\end{array}
\right ). 
\]
The constant $c$ appears because $K$ is
a conformal Killing vector.

Following the method of Finley and Pleba\'nski \cite{FP79} and using a freedom
in the heavenly potential $\Theta$ as well as in the choice of coordinates
we find the general solution to be
\[
K=(cZ + b)\p_Z+(cX-2bZ)\p_X.
\]

The conformal Killing equations  ${\cal L}_Kg=cg$ now yield
\[
{\cal L}_K(\Th_{XX})=-c\Th_{XX}, \qquad
{\cal L}_K(\Th_{XY})=b, \qquad
{\cal L}_K(\Th_{YY})=c\Th_{XX}.
\]
Let $U$ and $T$ be functions such that $K=\p_T$ and
${\cal L}_K(U)=0$. For $c\neq 0$ we can take
\[
T=\frac{\ln{(cZ+b)}}{c},\qquad U=\frac{2b}{c}T+\frac{2b^2+Xc^2}{c^2(cZ+b)}.
\]
The compatibility conditions for the Killing equations imply the
existence of $\t{G}=\t{G}(Y, W, U)$ such that
\[
\Th_{XX}=e^{-cT}\t{G}_{UU}, \qquad
\Th_{XY}=\t{G}_{YU}+bT, \qquad \Th_{YY}=e^{cT}\t{G}_{YY}.
\]
The heavenly equation (\ref{secondeq}) becomes
\[
bU+\frac{2b}{c}\t{G}_{YU}+c(\t{G}_Y-U\t{G}_{YU})+
\t{G}_{UW}+\t{G}_{YY}\t{G}_{UU}-\t{G}_{YU}^2=0.
\]
To obtain a simplified form define 
\[
G(Y, W, U)=\t{G}(Y, W, U)+\frac{b}{c}UY+\frac{b^2}{c^2}UW
\]
so that
\be
\label{form1}
bU+c({G}_Y-U{G}_{YU})+
{G}_{UW}+{G}_{YY}{G}_{UU}-{G}_{YU}^2=0.
\ee
Now rewrite (\ref{form1}) in terms of differential forms
\begin{eqnarray}
\label{form2}
&&bU\d Y\wedge\d U\wedge\d W+
c(G_Y\d Y\wedge\d U\wedge\d W-U\d G_U\wedge\d U\wedge\d W)\nonumber\\
&&+\d G_U\wedge \d Y\wedge\d U+\d G_Y\wedge\d G_U\wedge\d W
=0.
\end{eqnarray}
Define 
\[
x=G_U,\qquad y=Y
,\qquad t=-W,\qquad H(x, y, t)=xU(x, y, t) 
-G(Y, W, U(x, y, t)), 
\]
and preform a Legendre transform
\begin{eqnarray*}
\d H&=&\d (xU-G)=U\d x- G_Y\d Y- G_W\d W\\
&=&H_x\d x+ H_y\d y+H_t\d t.
\end{eqnarray*}
Therefore
\[
U=H_x, \qquad G_Y=-H_y, \qquad G_W=H_t.
\]
Differentiating these relations we find
\[
G_{UU}=\frac{1}{H_{xx}},\qquad
G_{YU}=-\frac{H_{xy}}{H_{xx}},\qquad
G_{YY}=-H_{yy}+\frac{H_{xy}^2}{H_{xx}}.
\]
The differential equation for $H(x, y, t)$ is obtained from (\ref{form2})
\be
\label{Heq1}
H_{xt}+ bH_xH_{xx}+c(H_{xy}H_x-H_yH_{xx})=H_{yy}.
\ee
Setting $u=H_x, w=-H_y$ 
we recover  (\ref{metric}), where $(u, w)$ solve (\ref{uw_eq}).

\end{document}